\documentstyle[epsfig]{article}

\newcounter{eqnletter}[equation]
\catcode`\@=11
\@addtoreset{equation}{section}
\catcode`\@=12

\begin{document}
 {\centerline {\LARGE {\bf  Real symmetric random matrices and replicas }}}
\vskip 1 cm
\centerline  {Giovanni M.Cicuta$^{(1)}$and Henri Orland$^{(2)}$ }
\vskip .5 cm
{\small \centerline  {$(1)$\,{\it Dip. di Fisica, Univ.di Parma, Parco Area delle Scienze 7A, 43100 Parma, Italy }}
\centerline  {{\it and INFN, Sez.di Milano, Gruppo di Parma}, email: cicuta@fis.unipr.it}
\centerline { $(2)$\,{\it Service de Physique Th\'eorique, CEA-Saclay, }}
\centerline {{\it 91191 Gif-sur-Yvette Cedex, France}, email: henri.orland@cea.fr}

\vskip .7 cm
{\centerline{\bf Abstract}}

Various ensembles of random matrices with independent entries are analyzed by the replica formalism in the large-$N$ limit. A result on the Laplacian random matrix with Wigner-rescaling is generalized to arbitrary probability distribution.\\
 
\section{Introduction}

Since the introduction of bosonic replicas in the study of disordered systems and random matrices several decades ago  \cite{sam}, most analytic investigations followed a path with three basic steps :\\
1) After defining a replica partition function, it is easy to
average it over the ensemble of random matrices to obtain a replica partition function where the interaction of the replica fields
depends on the probability distribution of the ensemble.\\
2) A set of auxiliary fields are introduced by Hubbard-Stratonovich 
identities to decouple the interacting pairs, transforming the replica partition function in a form suitable to the thermodynamic limit. In recent years a Gaussian functional identity conveniently replaced the set of auxiliary fields.\\
3) The saddle-point method leads to a non-linear integral equation of  very different complexity : it is trivially solved in the case of Wigner matrices whereas it is extremely difficult in the case of the Laplacian of a random graph.\\

This paper  follows the well established steps, respectively in sections 2, 3, 4, with greater generality.
Instead of choosing a definite probability distribution, we consider the off-diagonal matrix elements to be independent identically distributed (i.i.d.) random variables, where their probability distribution is coded by the cumulants. The diagonal matrix elements will be i.i.d. random variables statistically independent from the off-diagonal matrix elements or, as in the case of the Laplacian matrix, statistically dependent on them. In either case (statistically dependent or not) different specifications of the cumulants correspond to a full matrix or a sparse one.
This general approach does not lead to complications and all the three above mentioned steps are performed.\\
In sect.3 we discuss some puzzling features of the useful functional representation that decouples pairs of replica fields and its equivalence with the older method of auxiliary variables. The reader who is satisfied with formal inversion of operators without  closer inspection, may safely skip reading this section.\\
Sect.4 contains the saddle-point method , the non-linear integral equation and the new result of the paper : the limiting resolvent of the Laplacian of a random matrix {\bf with i.i.d. entries,  rescaled a-la-Wigner}, may be computed by a naive use of Pastur law of addition of random matrices, despite the statistically dependence of the diagonal entries. Our new result generalizes a previous one 
\cite{sta}, obtained 
 by diagrammatic methods and the assumption that the off-diagonal matrix entries are i.i.d. central Gaussian variables.\\

We think that the general and straightforward approach of this paper has some merits : indeed, as a by product, we obtain, at the beginning of Sect.4 the well known  Pastur law of addition of random matrices, in the replica formalism.\\

Sect.4 ends with the non-linear integral equation pertinent to the random graph. We included it to emphasize the generality of the replica formalism and the difference between Wigner-rescaled matrix ensembles (and their Laplacians) which leads to analytic solutions whereas ensembles where the matrix entries are of order $1/N$ are much more difficult.\\

It seems likely that everything in the paper may equally be expressed in terms of the supersymmetric method.\\

\section{Exact representations}

\noindent {\bf Notation.}\\
Let $H$ be a $N \times N$ real symmetric matrix. Let us suppose the diagonal entries $H_{rr}$ are $N$ i.i.d. random variables and the probability density of any of them is $ P_d(H_{rr})$. Let us suppose the off-diagonal entries $H_{r,s}$, where $r<s$, are $(N^2-N)/2$ i.i.d. random variables and the probability density of any of them is $P_{off}(H_{r,s})$. Let us write the expectations
$$ <d^m>=\int (H_{rr})^m\,P_d(H_{rr})\,dH_{rr} \qquad , \qquad
<v^m>=\int (H_{rs})^m\,P_{off}(H_{rs})\,dH_{rs} \qquad  \qquad $$

Let $G(z)$ be the  matrix resolvent
\begin{eqnarray}
 G_{ij}(z)=\left((zI_N-H)^{-1}\right)_{ij} \qquad , \qquad {\rm Im}\, z >0 \qquad
 \label{h.0}
\end{eqnarray}
Perturbation expansion in powers of $H$ shows that the expectations of all diagonal elements  $<G_{ii}(z)>$ are equal and of the off-diagonal elements $<G_{ij}(z)>$ are also equal. 
\begin{eqnarray}
<G_{ij}(z)>= \left\{ \begin{array}{cc}
 g(z) \qquad {\rm if} \; i=j & , \qquad \frac{1}{N}<{\rm tr}\,G(z)>=g(z)\\ 
g_1(z) \qquad {\rm if} \; i \neq j & \end{array} \right. \qquad \qquad
 \label{h.0a}
\end{eqnarray}
The representation of the  inverse matrix and usual definitions for the partition function $Z(z)$ and the averaged spectral density $\rho(\lambda)$ are

\begin{eqnarray}
&& G_{ij}= 
\frac{ (-i)\int_{-\infty}^{\infty}\left(\prod_{i=1}^N d\phi_i\right)\, e^{\frac{i}{2}\sum_{i,j}\phi_i(z-H)_{ij}\phi_j}\, \phi_i\phi_j}{ \int_{-\infty}^{\infty}\left(\prod_{i=1}^N d\phi_i\right)\, e^{\frac{i}{2}\sum_{i,j}\phi_i(z-H)_{ij}\phi_j} }\qquad ,
 \qquad \qquad \nonumber\\
&& 
 Z(z)\equiv \int_{-\infty}^{\infty}\left(\prod_{i=1}^N d\phi_i\right)\, e^{\frac{i}{2}\sum_{i,j}\phi_i(z-H)_{ij}\phi_j} =
\frac{ (2i\pi)^{N/2}}{\sqrt{\det (z I_N-H)}}
\quad , \quad \nonumber\\
&&g(z)=- \frac{2}{N}\frac{\partial}{\partial z}<\log Z(z)> \qquad ,\qquad 
\rho(\lambda)=-\frac{1}{\pi} {\rm Im} \, g(\lambda) \qquad \qquad
 \label{h.1}
\end{eqnarray}

 To get rid of the denominator of the resolvent, let us introduce $n$ copies $\phi_i^{(\alpha)}$ , $\alpha=1,2,..,n$

 \begin{eqnarray}
&&Z_n(z)=Z^n(z)= \int_{-\infty}^{\infty}\left(\prod_{\alpha=1}^n\prod_{i=1}^N d\phi_i^{(\alpha)}\right)\, e^{\frac{i}{2}\sum_{i,j,\alpha} \phi_i^{(\alpha)}(z-H)_{ij}\phi_j^{(\alpha)}} \qquad , \nonumber\\
&&g(z)=-\frac{2}{N}\frac{\partial}{\partial z} \lim_{n \to 0}<\frac{Z_n(z)-1}{n}>  \qquad \qquad 
 \label{h.1bb}
\end{eqnarray}

\noindent {\bf Exact representations.}\\
The above introduction of bosonic replicas allows an easy averaging over the matrix ensemble and leads us to an
 exact representation for any $N$ for the averaged $<Z_n(z)>$ and resolvent, see eq. (\ref{c.2}) below, in terms of cumulants of the probability distribution of the matrix entries.\\

 Let us assume we may first average on the entries $H_{ij}$ , then integrate over $\phi_r^{(\alpha)}$, finally take the limit $n\to 0$. Let us split the diagonal from the off-diagonal entries
  \begin{eqnarray}
&& <e^{\frac{i}{2}\sum_{i,j,\alpha}\phi_i^{(\alpha)}(z-H)_{ij}\phi_j^{(\alpha)}}>=\qquad \nonumber\\
 &&\quad e^{\frac{i}{2}z\sum_{i,\alpha}(\phi_i^{(\alpha)})^2 } 
 \prod_{i=1}^N
<e^{-\frac{i}{2}H_{ii} \sum_{\alpha}(\phi_i^{(\alpha)})^2 }>\prod_{r<s}<e^{-i H_{rs} \sum_{\alpha}\phi_r^{(\alpha)}\phi_s^{(\alpha)} }>
\qquad \qquad
  \label{h.1d}
\end{eqnarray}

 This has a compact representation in terms of cumulants of the moments of the matrix entries.
Let $x$ be a random variable, we may define the cumulants $c_j$ by equating the powers of $\alpha$ of the exponential generating functions
 \begin{eqnarray}
&&\log <e^{\alpha \,x}>=\log \left[\sum_{k=0}^{\infty}\frac{\alpha^k}{k!}<x^k>\right]=\sum_{j=1}\frac{1}{j!} \alpha^j \,c_j \qquad , \qquad {\rm that}\qquad {\rm is}\qquad \qquad \nonumber\\
&&c_1=<x> \quad , \quad c_2=<x^2>-<x>^2 \quad , \quad  c_3=<x^3>-3<x^2><x>+2<x>^3 \qquad ,\nonumber\\
&&c_4=<x^4>-4<x^3><x>-3<x^2>^2+12<x^2><x>^2-6<x>^4\quad , \cdots \qquad \nonumber\\
 \label{c.1}
\end{eqnarray}

To simplify the notation, we call $c_{k,d}$, $c_{k,v}$ the k-cumulants of the diagonal and off-diagonal entries of the random matrix $H$.\\
 
One obtains the exact representation \footnote{ It is easy to check that this integral representation is an absolute convergent integral for every integer $n$. Indeed $ \left|{\rm exp} \left\{ \sum_{j=1} \frac{1}{j!} c_{j,d}\sum_{r=1}^N
\left(-\frac{i}{2}\sum_\alpha   \phi_r^{(\alpha)} \phi_r^{(\alpha)}
 \right)^j\right\}\right|=\prod_{r=1}^N\left|<e^{-\frac{i}{2}H_{rr} \sum_{\alpha}(\phi_r^{(\alpha)})^2 }>\right|\leq 1$ and similar relation holds for the factor containing $H_{r,s}$.}

 \begin{eqnarray}
&&<Z_n(z)>= \int \prod_{i,\alpha} d\phi_i^{(\alpha)} \, e^{\frac{i}{2}z\sum_{r=1}^N\sum_{\alpha=1}^n \phi_r^{(\alpha)}\phi_r^{(\alpha)}  }\quad e^F \qquad , \qquad \nonumber\\
&&F= 
 \sum_{j=1} \frac{1}{j!}\Bigg[ c_{j,d}\sum_{r=1}^N
\left(-\frac{i}{2}\sum_\alpha \left(\phi_r^{(\alpha)}\right)^2 \right)^j
 +c_{j,v} \sum_{1 \leq r<s \leq N}
\left(-i\sum_\alpha \phi_r^{(\alpha)}\phi_s^{(\alpha)} \right)^j\Bigg] \qquad \qquad
\label{c.2}
\end{eqnarray}
 
\noindent {\bf The random Laplacian  matrix}\\    
An analogous integral representation holds for a random Laplacian matrix.\\
Let us suppose that the real symmetric random matrix $H_{ij}$ is a Laplacian matrix: we assume that the diagonal elements are $H_{ii}\stackrel{\mathrm{def}}{=}
-\sum_{j \neq i}H_{ij}$. Again we assume that the $N(N-1)/2$ random variables $H_{ij}$ with $i<j$ are i.i.d.\\
Since
 \begin{eqnarray}
\sum_{i,j}\phi_i H_{ij}\phi_j&=&-\sum_i (\phi_i)^2 \left(\sum_{j<i}H_{ij}+\sum_{j>i}H_{ij}\right)+2\sum_{i<j}\phi_i \phi_j H_{ij}= \qquad \nonumber\\
&=&- \sum_{i<j}H_{ij}\left(\phi_i-\phi_j\right)^2 \qquad \qquad
\label{la.1}
\end{eqnarray}
the partition function , the averaged resolvent and their replicated versions are
 \begin{eqnarray}
 &&Z(z)\equiv \int_{-\infty}^{\infty}\left(\prod_{i=1}^N d\phi_i\right)\, e^{\frac{i}{2}z\sum_{i}(\phi_i)^2} 
\, e^{\frac{i}{2}\sum_{i<j}H_{ij}\left(\phi_i-\phi_j\right)^2 } 
\quad , \quad \nonumber\\
&&g(z)=- \frac{2}{N}\frac{\partial}{\partial z}<\log Z(z)> \qquad ,\qquad 
 \nonumber\\
&&Z_n(z)=Z^n(z)= \int_{-\infty}^{\infty}\left(\prod_{\alpha=1}^n\prod_{i=1}^N d\phi_i^{(\alpha)}\right)\,  
 e^{ \frac{i}{2}z\sum_{i,\alpha}\left(\phi_i^\alpha\right)^2} 
\, e^{\frac{i}{2}\sum_{i<j}H_{ij}\sum_\alpha\left(\phi_i^\alpha-\phi_j^\alpha\right)^2 } 
 \quad , \quad \nonumber\\
&&g(z)=-\frac{2}{N}\frac{\partial}{\partial z} \lim_{n \to 0}<\frac{Z_n(z)-1}{n}>  \qquad \qquad 
 \label{la.3}
\end{eqnarray}
Again one may perform the average over the matrix ensemble to obtain an exact  expression in terms of the cumulants $c_j$ of the moments of the off-diagonal entries
\begin{eqnarray}
&&<Z_n(z)>=\int \prod_{i,\alpha} d\phi_i^{(\alpha)} \, e^{\frac{i}{2}z\sum_{i=1}^N\sum_{\alpha=1}^n \phi_i^{(\alpha)}\phi_i^{(\alpha)}  }\quad e^F \qquad , \qquad \nonumber\\
&&F= 
 \sum_{1 \leq r<s \leq N}\sum_{j=1}\frac{1}{j!}c_j\left[\frac{i}{2}\sum_\alpha \left(\phi_r^\alpha -\phi_s^\alpha \right)^2\right]^j
\qquad \qquad 
 \label{la.4}
\end{eqnarray}

\noindent {\bf Special cases.}\\
It may be useful to recover from the general representations in eqs.(\ref{c.2}) and (\ref{la.4}) some cases corresponding to special probability distribution for random matrices, which have been studied by use of bosonic replicas.\\

\noindent
{\bf $1.$ The Gaussian ensemble}. In the Gaussian orthogonal ensemble (GOE) the joint probability density of the real symmetric random matrix is $P(H)$
$$P(H)=cons. \times e^{-\frac{1}{2\sigma^2}{\rm tr}\,H^2}\qquad $$
Then all cumulants are zero except the second one, $c_{2,d}=\sigma^2$ and $c_{2,v}=\sigma^2/2$.\\
The exponent $F$ in eq.(\ref{c.2}) is
\begin{eqnarray}
F= -\frac{\sigma^2}{8}\sum_{r,s} \left[\sum_\alpha 
\phi_r^{(\alpha)}\phi_s^{(\alpha)}\right]^2=  
 -\frac{\sigma^2}{8}\sum_{\alpha , \beta}\left[\sum_r
 \phi_r^{(\alpha)}\phi_r^{(\beta)}\right]^2 \qquad 
\qquad \qquad 
 \label{la.4a}
\end{eqnarray}

 The $F$ exponent of the replicated partition function
of the Laplacian matrix with Gaussian entries (\ref{la.4}) is
\begin{eqnarray}
F=-\frac{\sigma^2}{16}
  \sum_{1 \leq r<s \leq N}
\left[\sum_\alpha \left(\phi_r^{(\alpha)} -\phi_s^{(\alpha)} \right)^2\right]^2
\qquad \qquad 
 \label{la.4b}
\end{eqnarray}

\noindent
{\bf $2.$ The sparse random matrix.} In the case of the sparse random matrix considered by Rodgers and Bray \cite{rod} , the independent entries  $J_{ij}$, $i 
\leq j$,  have a even probability distribution so that
$<v^{2k+1}>=0$, $<v^{2k}>=p/N$ for $k \geq 1$. The same even probability distribution and its asymmetric generalization have been studied in \cite{seme}.\\
The equation to evaluate the cumulants is
$$\log \left[<e^{-i\alpha x}>\right]=\log \left[1+\frac{p}{N}\left(-1+\cos \,\alpha\right)\right]= \sum_{j=1}\frac{c_j}{j!}(-i\alpha)^j\qquad \qquad$$
Then the exponent $F$ in eq.(\ref{c.2}) is
 \begin{eqnarray}
F&=&\sum_r \log \left[1+\frac{p}{N}\left(-1+\cos \frac{1}{2}\sum_\alpha    
(\phi_r^{(\alpha)})^2 \right)\right]+ \qquad \qquad\nonumber\\
&+&\sum_{1\leq r < s \leq N}\log \left[1+\frac{p}{N}\left(-1+\cos \sum_\alpha \phi_r^{(\alpha)}\phi_s^{(\alpha)}\right)\right]
\qquad \qquad
 \label{la.4bb}
\end{eqnarray}
If only the limit $N \to \infty$ is investigated, the first term in the above equation may be dropped and the second one may be approximated
 \begin{eqnarray}
F\sim
 \frac{1}{2}\sum_{r,s}\log \left[1+\frac{p}{N}\left( -1+\cos \sum_{\alpha=1}^n\phi_r^{(\alpha)} \phi_s^{(\alpha)}\right)\right]\sim
\frac{p}{2N}\sum_{r,s}\left(-1+
\cos \sum_{\alpha=1}^n\phi_r^{(\alpha)} \phi_s^{(\alpha)}\right)
\qquad \qquad
\label{c.4b}
\end{eqnarray}

\noindent
{\bf $3.$ The random graph}. The case of the random graph was often analyzed as a Bernoulli probability distribution of the off-diagonal 
entries of the random matrix 
$H_{ij},\  i<j$, where $H_{ij}$ is $-1/q$ with probability $q/N$ and $0$ with probability $1-q/N$ , while the diagonal entries  vanish , $H_{jj}=0$.\\ Then $<v^k>=\frac{q}{N}\left(-1/q\right)^k$ for $k\geq 1$ and 
the equation to evaluate the cumulants $c_j$ is
 \begin{eqnarray}
\log <e^{\alpha x}>=\log \left[1+\frac{q}{N}\left(e^{-\frac{\alpha}{q}}-1\right)\right]=
\sum_{j=1}\frac{c_j}{j!}\alpha^j\qquad \qquad
\label{c.4bc}
\end{eqnarray}
Then the exponent $F$ in eq.(\ref{c.2}) is
\begin{eqnarray}
 F=
 \sum_{r<s}\log \left[1-\frac{q}{N}+\frac{q}{N}e^{-\frac{1}{q}\sum_{\alpha=1}^n\phi_r^\alpha \phi_s^\alpha}\right] \qquad \qquad
\label{c.4a}
\end{eqnarray}
The $F$ exponent of the replicated partition function
of the Laplacian matrix (\ref{la.4}) is
\begin{eqnarray}
 F=\sum_{1 \leq r<s\leq N}\log \left[1-\frac{q}{N}+\frac{q}{N}e^{-\frac{i}{2q}\sum_\alpha ( \phi_r^{(\alpha)}-\phi_s^{(\alpha)})^2}\right]\qquad \qquad
\label{c.4ab}
\end{eqnarray}

The spectral density corresponding to eq.(\ref{c.4ab}) is studied in \cite{for}, \cite{bir}, \cite{dea}.\\

\noindent
{\bf $4.$ The diagonal random matrix.} For future reference we also recover the  partition function and resolvent  for the simple case of the diagonal random matrix. Of course in this trivial case, replicas are totally unnecessary.\\

From eq.(\ref{c.1}) and (\ref{c.2})
\begin{eqnarray}
&&<e^{\alpha \, d}>=\sum_{k=0}^{\infty}\frac{\alpha^k}{k!}<d^k>=e^{\sum_{j=1}^{\infty}\frac{1}{j!}\alpha^j c_{j,d} }
\qquad , \qquad \qquad \nonumber\\
&&<Z_n(z)>= \int \prod_{i,\alpha}d\phi_i^{(\alpha)}\,
e^{\frac{i}{2}z \sum_{r=1}^N {\vec \phi}_r\cdot {\vec \phi}_r}\prod_{r=1}^N\sum_{k=0}^{\infty}\frac{1}{k!}<d^k>\left( \frac{-i}{2}({\vec \phi}_r \cdot {\vec \phi}_r)\right)^k=
\qquad \qquad \nonumber\\
&&\qquad =\left[
\int \prod_{\alpha=1}^n d\phi^{(\alpha)}\,
e^{\frac{i}{2}z  {\vec \phi}\cdot {\vec \phi}}
\sum_{k=0}^{\infty}\frac{1}{k!}<d^k>\left( \frac{-i}{2}({\vec \phi}\cdot {\vec \phi})\right)^k \right]^N=\nonumber\\
&&\qquad=\left[ \sum_{k=0}^{\infty}\frac{(-1)^k}{k!}<d^k>\left(\frac{\partial}{\partial z}\right)^k
\int \prod_{\alpha=1}^n d\phi^{(\alpha)}\,
e^{\frac{i}{2}z  {\vec \phi}\cdot {\vec \phi}}
\right]^N=\nonumber\\
&&\qquad = \left[e^{in\pi/4}(2\pi)^{n/2}
\sum_{k=0}^{\infty}\frac{1}{k!}<d^k>
\frac{n}{2}\left(\frac{n}{2}+1\right)\cdots\left(\frac{n}{2}+k-1\right)z^{-\frac{n}{2}-k}\right]^N\qquad
\label{d.1}
\end{eqnarray}

Then for $n \to 0$
\begin{eqnarray}
&&<Z_n(z)> \sim \left[z^{-n/2}+\frac{n}{2} \sum_{k=1}^{\infty}\frac{1}{k}<d^k>
z^{-k}\right]^N\qquad ,\qquad \nonumber\\
&&g(z)=\sum_{k=0}^{\infty}\frac{<d^k>}{z^{k+1}}=\int d\lambda \frac{\rho(\lambda)}{z-\lambda} \qquad \qquad
\label{d.2}
\end{eqnarray}

\section{Hilbert-Schmidt with replicas}
The next step in the analysis of the replicated partition function , eqs.(\ref{c.2}) and (\ref{la.4}), is to transform the part of the integrand of $<Z_n(z)>$ that involves sums on pairs of sites into a form involving only sums over single sites. Having done that, the partition function may be written as  an integral  where the number of sites $N$ appears just as a parameter.\\
This was done by Rodgers and Bray \cite{rod} by introducing a set of auxiliary variables $q_\alpha , q_{\alpha \beta} , q_{\alpha \beta \gamma} ,..$ with the corresponding Hubbard-Stratonovich identities. The same authors used a functional representation in the case of a Laplacian matrix, \cite{bra}. \\
It seems very convenient to use a functional  Hubbard-Stratonovich transform
\begin{eqnarray}
e^{-\frac{1}{2}\sum_{r,s} K(\phi_r,\phi_s)}=
\int {\cal D}g \,e^{-\frac{1}{2} \int d\phi_a \,d\phi_b \,g(\phi_a)C(\phi_a,\phi_b)\,g(\phi_b)-i\sum_{j=1}^N g(\phi_j)} \qquad
\label{p.1}
\end{eqnarray}
suggested by Fyodorov and Mirlin \cite{fio1} and discussed in more detail in the appendix of \cite{fio2}.\\
An analogous functional integral representation is discussed by Parisi \cite{par} for euclidean random matrices
\begin{eqnarray}
e^{-\frac{\beta}{2}\sum_{r,s}V(x_r-x_s)}=(const.)\int d\sigma\,
e^{\frac{1}{2\beta}\int dx \, dy\sigma(x)V^{-1}(x-y)\sigma(y)+\sum_j \sigma(x_j)}
\label{p.2}
\end{eqnarray}
As it is indicated by Parisi, eq.(\ref{p.2}) follows from the familiar identity
\begin{eqnarray}
(const.)\int d\sigma\,
e^{\frac{1}{2\beta}\int dx \, dy\sigma(x)V^{-1}(x-y)\sigma(y)+\int dx J(x)\sigma(x)}=e^{-\beta\int dx \, dy \,J(x)V(x-y)J(y)}
\nonumber\\
\label{p.3}
\end{eqnarray}
after the choice of the source $J(x)=\sum_{j=1}^N \delta(x-x_j)$.\\

 \noindent {\bf The problem}\\
In the Gaussian case, the sum over pairs of sites, occurring in the exponent $F$, eqs.(\ref{c.2}) and (\ref{la.4}), 
 is a homogeneous polynomial of degree $4$. In the case of a general probability distribution,  cumulants of arbitrary high order are not zero. Yet it
 will be apparent in the next section that also in this case the large-$N$ saddle point analysis of  the resolvent leads to the Gaussian case in leading order : cumulants $c_{j,v}$ with $j \geq 3$ are associated to non-leading terms, if the random matrix elements are rescaled a-la-Wigner.\\
Then these integral representations are rather puzzling because they involve the inverse of the integral operator $K(\phi_r,\phi_s)$ which is  a polynomial (in the most important case) and therefore the eigenspace corresponding to the null eigenvalue has infinite dimension.\\
 Still the method is very useful and leads to correct results. It was correctly indicated by Fyodorov how to evaluate the inverse operator in a subspace. This section is a little elaboration on Fyodorov appendix \cite{fio2}.\\ Our conclusion is that, provided a regularization of the problem is allowed, the 
functional  Hubbard-Stratonovich transform is correct and identical to the set of usual Hubbard-Stratonovich identities.\\

\noindent {\bf The symmetric matrix}\\
Let us consider all replicas in a box , $-L \leq \phi_r^{(\alpha)} \leq L$, $r=1,..,N$, $\alpha=1,..,n$.\\
 Let ${\vec \phi_r}$ be the $n$-component vector with components $\phi_r^{(\alpha)}$, $\alpha=1,..,n$.\\
Consider the integral operator kernel $K({\vec \phi_r }, {\vec \phi_s})$ with finite rank $D$ (to make the discussion simpler, we shall ignore the imaginary unit factors in this section)
\begin{eqnarray}
&& K({\vec \phi_r },{\vec \phi_s})= \sum_{j=1}^D\frac{c_{j}}{j!}\left(\sum_{\alpha=1}^n \phi_r^{(\alpha)}\phi_s^{(\alpha)}\right)^j=
\sum_{j=1}^D\frac{c_{j}}{j!}\left( {\vec \phi_r}\cdot {\vec \phi_s}\right)^j =\qquad \qquad \nonumber\\
&&=\sum_{j=1}^D\frac{c_{j}}{j!} \sum_{\alpha_1,..,\alpha_j}
\phi_r^{(\alpha_1)}\cdots\phi_r^{(\alpha_j)}\phi_s^{(\alpha_1)}\cdots
\phi_s^{(\alpha_j)}
\qquad ,\qquad \nonumber\\
&&\sum_{r,s}K({\vec \phi_r },{\vec \phi_s})=  \sum_{j=1}^D\frac{c_{j}}{j!} \sum_{\alpha_1,..,\alpha_j}
\left(\sum_{r=1}^N \phi_r^{(\alpha_1)}\cdots\phi_r^{(\alpha_j)}\right)^2 \qquad \qquad \qquad
\label{fa.20}
\end{eqnarray}
It maps an arbitrary integrable function $f({\vec \phi_s})$ 
to the function $g({\vec \phi_r}) \in S_D$ which is the linear finite $D$-dimensional vector space
 spanned by the set $s$ of linearly independent vectors
\begin{eqnarray}
&& s=\left\{  \phi_r^{(\alpha_1)} \, , \, \phi_r^{(\alpha_1)}\phi_r^{(\alpha_2)}  \, ,..,\,\phi_r^{(\alpha_1)} \phi_r^{(\alpha_2)}..\phi_r^{(\alpha_D)}   \right\} =\left\{ b_{\nu}({\vec \phi_r })\right\}
 \quad ,\quad \nu=1,\cdots ,D \quad , \nonumber\\
&&
\int_{-L}^L ..\int_{-L}^L K( {\vec \phi_r },{\vec \phi_s}) \,
f({\vec \phi_s}) d{\vec \phi_s}=g({\vec \phi_r})
\qquad ,\qquad   d{\vec \phi_s}=\prod_\alpha^n d\phi_s^{(\alpha)} \qquad \qquad
\label{fa.21}
\end{eqnarray}
The vector bases $s$ is made of $D$ non-orthogonal vectors
$b_{\nu}({\vec \phi_r })$ .
 The integral operator $K({\vec \phi_r }, {\vec \phi_s})$ has $D$ orthonormal eigenvectors $e_\nu$, corresponding to non-zero eigenvalues
\begin{eqnarray}
&&\int_{-L}^L ..\int_{-L}^L K({\vec \phi_r }, {\vec \phi_s}) \,
e_{\nu}({\vec \phi_s}) \,d{\vec \phi}_s=\lambda_\nu e_{\nu}({\vec \phi}_r) \qquad , \qquad \nonumber\\
&& K({\vec \phi_r },{\vec \phi_s})= \sum_{\nu=1}^D \lambda_{\nu}e_{\nu}({\vec \phi_r}) e_{\nu}({\vec \phi_s}) 
 \qquad , \qquad C({\vec \phi}_r ,{\vec \phi}_s) \stackrel{\mathrm{def}}{=}
\sum_{\mu=1}^D \frac{1}{\lambda_\mu}e_\mu ({\vec \phi}_r )e_\mu ({\vec \phi}_s) \quad , \nonumber\\
&&\int_{-L}^L K({\vec \phi_r },{\vec \phi_s})C( {\vec \phi_s}, {\vec \phi_t})\,d {\vec \phi_s}=
\sum_{\nu=1}^D e_{\nu} ({\vec \phi_r}) e_{\nu}(  {\vec \phi_t})=I_D
({\vec \phi_r},{\vec \phi_t}) 
\qquad \qquad
\nonumber\\
\label{fa.22}
\end{eqnarray}
The vector bases $\left\{ b_{\nu} ( {\vec \phi_r } )\right\}$ , and the orthonormal bases $\left\{ e_{\nu}({\vec \phi_r })\right\}$ are related by a linear invertible matrix $A$
\begin{eqnarray}
&&b_\nu=\sum_{\mu} A_{\nu \mu}e_{\mu} \quad  , \quad
e_\nu=\sum_{\mu} \left(A^{-1}\right)_{\nu \mu}b_{\mu} \quad  , \quad
A_{\nu \mu}= \int_{-L}^L     b_\nu({\vec \phi_r }) e_\mu({\vec \phi_r })d{\vec \phi_r }
\qquad \qquad \qquad 
\label{fa.23}
\end{eqnarray}

As indicated by Fyodorov, the functional Hubbard-Stratonovich transform (\ref{p.1}) is a compact representation of the multiple integral \footnote{In the functional integrals, ${\cal D}g$ refers to the product, properly normalized, of the coordinates of the function $g({\vec \phi})$ with respect to an orthogonal bases, then ${\cal D}g=\prod_\nu \sqrt{\lambda_\nu/ 2 \pi} dz_\nu$ and eq.(\ref{p.1}) should be written
\begin{eqnarray}
e^{-\frac{1}{2}\sum_{r,s} K({\vec \phi}_r ,{\vec \phi}_s)}= \sqrt{ \det C} 
\int {\cal D}g \,e^{-\frac{1}{2} \int d{\vec\phi}_a \,d{\vec \phi}_b \,g({\vec \phi}_a)C({\vec \phi}_a,{\vec \phi}_b)\,g({\vec \phi}_b)-i\sum_{j=1}^N g({\vec\phi}_j)} \qquad \qquad
\label{p.1b}
\end{eqnarray}
}
\begin{eqnarray}
&&e^{-\frac{1}{2}\sum_{r,s} K({\vec \phi}_r ,{\vec \phi}_s)}= \prod_{\nu=1}^D
e^{-\frac{1}{2}\lambda_\nu \left[\sum_{r=1}^N e_\nu({\vec \phi}_r )\right]^2}=
\prod_{\nu=1}^D
\int_{-\infty}^{\infty} 
\frac{ dz_\nu}{\sqrt{  2\pi}} 
 \,e^{-\frac{1}{2}z_{\nu}^2
 -i\sum_{r=1}^N \sqrt{\lambda_\nu}z_\nu e_\nu({\vec \phi}_r)} \qquad ,
\nonumber\\
&& 
g({\vec \phi}_r )=\sum_{\nu=1}^D \sqrt{\lambda_\nu}z_\nu e_\nu({\vec \phi}_r )
\quad , \quad  
\int d{\vec \phi}_a \,d{\vec \phi}_b \,g({\vec \phi}_a)C({\vec\phi}_a,{\vec \phi}_b)\,g({\vec\phi}_b)=\sum_\mu z_{\mu}^2 \qquad \qquad
\label{fa.25}
\end{eqnarray}

After a trivial relabelling of the coefficients in the form (\ref{fa.20}), one uses  the usual set of Hubbard-Stratonovich identities
\begin{eqnarray}
e^{-\frac{1}{2}\sum_{r,s} K({\vec \phi}_r,{\vec\phi}_s)}=
e^{-\frac{1}{2}\sum_{\nu=1}^D c(\nu)\left(\sum_{r=1}^Nb_{\nu}({\vec \phi}_r) \right)^2}=
\prod_{\nu=1}^D  \int_{-\infty}^{\infty} 
\frac{dq_\nu}{\sqrt{2\pi}}  \,e^{-\frac{1}{2}q_{\nu}^2
 -i\sum_{r=1}^N \sqrt{c_\nu}q_\nu b_\nu({\vec \phi}_r)} \qquad
\label{fa.26}
\end{eqnarray}

Finally one may notice that the set of integration variables $q_\nu$ and $z_\nu$ are related by an orthogonal matrix ${\cal A}$ which shows the direct equivalence of the two multiple integrals (\ref{fa.25}) and (\ref{fa.26})
\begin{eqnarray}
&&q_\nu=\sum_{\mu=1}^D{\cal A}_{\nu \mu }z_\mu  \quad , \quad 
{\cal A}_{\nu \mu}=\sqrt{ \frac{c(\nu)}{\lambda_\mu}}A_{\nu \mu}
 \quad , \quad \nonumber\\
&&\sum_{\nu=1}^D {\cal A}_{\nu \tau}{\cal A}_{\nu \sigma}=\delta_{\tau \sigma}  \quad , \quad 
\sum_\nu q_\nu^2=\sum_\nu z_\nu^2 \quad , \quad \prod_{\nu=1}^D dq_\nu=\prod_{\nu=1}^D dz_\nu  
 \qquad  \qquad 
\label{fa.27}
\end{eqnarray}

The orthogonality of the matrix ${\cal A}$ (last line above) follows  from the equation
\begin{eqnarray}
\lambda_{\sigma}\delta_{\sigma \tau}=\sum_\nu c(\nu)A_{\nu \tau}A_{\nu \sigma} \qquad  \qquad 
\label{fa.28}
\end{eqnarray}
 which is obtained from
\begin{eqnarray}
K({\vec \phi_r },{\vec \phi_s})= \sum_\nu \lambda_\nu e_\nu({\vec \phi}_r)e_\nu({\vec \phi}_s)=\sum_\nu c(\nu)b_{\nu}({\vec \phi}_r)b_{\nu}({\vec \phi}_s) \qquad  \qquad 
\label{fa.29}
\end{eqnarray}
after multiplying and integrating $\int d{\vec \phi}_r d{\vec \phi}_s \,e_\sigma({\vec \phi}_r)\,e_\tau({\vec \phi}_s)$\\
Finally
\begin{eqnarray}
g({\vec \phi}_r)&=&\sum_{\nu=1}^D \sqrt{\lambda_\nu} z_\nu e_\nu({\vec \phi}_r)= \sum_{\nu,\sigma, \mu}
\sqrt{\lambda_\nu} \left[ ({\cal A}^{-1})_{\nu \sigma}q_\sigma \right]\left[(A^{-1})_{\nu \mu}b_\mu({\vec \phi}_r)\right]=
\nonumber\\
&&=\sum_{\nu,\sigma, \mu}
\sqrt{\lambda_\nu}  {\cal A}_{\sigma \nu}q_\sigma (A^{-1})_{\nu \mu}b_\mu({\vec \phi}_r)=
 \sum_\mu \sqrt{c_\mu} q_\mu b_\mu({\vec \phi}_r)
\qquad \qquad
\label{fa.30}
\end{eqnarray}
\noindent {\bf   The random Laplacian matrix}.\\
The replicated partition function for the symmetric random Laplacian matrix is (\ref{la.4}). Again ignoring  imaginary unit factors, the 
finite rank kernel of the integral operator is $K({\vec \phi}_r ,{\vec \phi}_s)$
\begin{eqnarray}
K({\vec \phi}_r ,{\vec \phi}_s)=\sum_{j=1}^D \frac{c_j}{2^jj!} \left[ \left( {\vec \phi}_r-{\vec \phi}_s\right)\cdot
\left( {\vec \phi}_r-{\vec \phi}_s\right)\right]^j
\qquad \qquad
\label{fa.40}
\end{eqnarray}
Let us assume the box regularization, as in the discussion of the symmetric random matrix above. Just as before, the kernel $K({\vec \phi}_r ,{\vec \phi}_s)$ in eq. (\ref{fa.40}) is a symmetric bounded integral operator that maps the integrable 
function $f({\vec \phi_s})$ 
to the function $g({\vec \phi_r}) \in S_D$, spanned by the non-orthogonal bases $b_\nu({\vec \phi}_r)$, see eq.
(\ref{fa.21}).\\
It is then possible to represent the kernel $K({\vec \phi}_r ,{\vec \phi}_s)$ in eq. (\ref{fa.40}) with its orthonormal bases and obtain Fyodorov functional Hubbard-Stratonovich representation or represent it on the non-orthogonal bases 
$b_\nu({\vec \phi}_r)$ , perform the sum over couples of sites and use the Hubbard-Stratonovich identities.\\

Let us consider the first term in the sum in eq.(\ref{fa.40}).
By the trivial identity
$$ {\vec \phi}_r \cdot {\vec \phi}_r+ {\vec \phi}_s \cdot {\vec \phi}_s= \left(1+{\vec \phi}_r \cdot {\vec \phi}_r\right)\left(1+{\vec \phi}_s \cdot {\vec \phi}_s\right)-\left({\vec \phi}_r \cdot {\vec \phi}_r\right)\left( {\vec \phi}_s \cdot {\vec \phi}_s\right)-1$$
The first term is rewritten as sum of factorized and symmetrical terms
\begin{eqnarray}
&& \frac{c_1}{2}\left( {\vec \phi}_r-{\vec \phi}_s\right)\cdot
\left( {\vec \phi}_r-{\vec \phi}_s\right)= \qquad \qquad \nonumber\\
&&=\frac{c_1}{2}\left\{
\left(1+{\vec \phi}_r \cdot {\vec \phi}_r\right)\left(1+{\vec \phi}_s \cdot {\vec \phi}_s\right)-\left({\vec \phi}_r \cdot {\vec \phi}_r\right)\left( {\vec \phi}_s \cdot {\vec \phi}_s\right)-1-2\left(
{\vec \phi}_r \cdot {\vec \phi}_s\right) \right\}
\qquad \qquad
\label{fa.41a}
\end{eqnarray}
All further terms in the sum are integer powers of the first term, then a finite sum  of factorized and symmetrical terms is obtained.\\

In particular, in the Gaussian case, where only the cumulant $c_2 \neq 0$, we have
\begin{eqnarray}
&&\sum_{r,s}K( {\vec \phi}_r ,{\vec \phi}_s)= \frac{c_2}{8} \sum_{r,s}\left[\left( {\vec \phi}_r-{\vec \phi}_s\right)\cdot
\left( {\vec \phi}_r-{\vec \phi}_s\right)\right]^2=\nonumber\\
&=&\frac{c_2}{8} \sum_{r,s}\left( {\vec \phi}_r \cdot {\vec \phi}_r +
{\vec \phi}_s \cdot {\vec \phi}_s\right)^2+4 \left(  
{\vec \phi}_r \cdot {\vec \phi}_s \right)^2-2 
\left(  
{\vec \phi}_r \cdot {\vec \phi}_s \right)\left(
{\vec \phi}_r \cdot {\vec \phi}_r+ {\vec \phi}_s \cdot {\vec \phi}_s
\right)=\qquad \nonumber\\
&=&\frac{c_2}{8} \left\{ \left[\sum_r(1+{\vec \phi}_r \cdot {\vec \phi}_r)^2 \right]^2+\left[\sum_r ({\vec \phi}_r \cdot {\vec \phi}_r)^2\right]^2+ 
N^2-2\left[\sum_r \left(1+ {\vec \phi}_r \cdot {\vec \phi}_r\right)\right]^2
\right.\nonumber\\
&& -2 \left[\sum_r {\vec \phi}_r \cdot {\vec \phi}_r \right]^2-2\left[ \sum_r ({\vec \phi}_r \cdot {\vec \phi}_r)(1+
{\vec \phi}_r \cdot {\vec \phi}_r) \right]^2+
4\sum_{\alpha \beta}\left[ \sum_r \phi_r^{(\alpha)}\phi_r^{(\beta)}\right]^2
\nonumber\\
&&  \left.-4\sum_\alpha \left[\sum_r \phi_r^{(\alpha)}(1+
{\vec \phi}_r \cdot {\vec \phi}_r )\right]^2+4\sum_\alpha \left[
\sum_r \phi_r^{(\alpha)}({\vec \phi}_r \cdot {\vec \phi}_r )\right]^2 
+4\sum_\alpha \left[\sum_r \phi_r^{(\alpha)}\right]^2\right\}
\qquad \qquad
\label{fa.41}
\end{eqnarray}
This form is suitable to linearize the squared sums of single sites
by the introduction of $9$ auxiliary variables   through Hubbard-Stratonovic identities. The method also works for arbitrary power $j>2$ of the basic form in eq.(\ref{fa.41a}), but it is cumbersome and by far less efficient than the functional integral representation.\\

\section{Large N}
As a preparation for the new result of this paper, about the limiting resolvent of Wigner-rescaled random Laplacians, we first consider an ensemble of Wigner-rescaled matrices which leads to a derivation of Pastur addition law of random matrices in the replica formalism.\\

\noindent {\bf Pastur addition law.}\\
The first problem here considered is the large $N$ behaviour for the symmetric matrix where the probability distribution of the off-diagonal entries is central Gaussian with Wigner scaling, while 
the probability distribution of the  diagonal entries is arbitrary and not rescaled. Then $c_{j,v}=0$ iff $j \neq 2$ and $c_{2,v} \to c_2/N$.\\
After separating the sums over single sites from sums over couples of sites, the replicated partition function (\ref{c.2}) becomes

 \begin{eqnarray}
\!\!&\!\!&\!<Z_n(z)>=\! \int \prod_{i,\alpha} d\phi_i^{(\alpha)} \, e^{\frac{i}{2}z\sum_{r}({\vec \phi}_r \cdot {\vec\phi}_r) } \quad e^{F_1}\, e^{\frac{c_2}{4N}\sum_r ({\vec \phi}_r \cdot {\vec \phi}_r)^2} e^{-\frac{c_2}{4N}
 \sum_{r,s }
\left(\sum_\alpha \phi_r^{(\alpha)}\phi_s^{(\alpha)} \right)^2}
\qquad , \qquad \nonumber\\
&&F_1\stackrel{\mathrm{def}}{=} 
 \sum_{j=1}\frac{1}{j!} c_{j,d}\sum_{r=1}^N
\left(-\frac{i}{2}\sum_\alpha \left(\phi_r^{(\alpha)}\right)^2 \right)^j
 \qquad \qquad \qquad
\label{hen.1a}
\end{eqnarray}
By using the functional representation (\ref{p.1b}) we have
 \begin{eqnarray}
<Z_n(z)>&=& \sqrt{ \det C} 
 \int {\cal D}g \,e^{-\frac{1}{2} (g,Cg)+N \log \,I[g]}
 \qquad ,\qquad  \qquad \nonumber\\
I[g] & \stackrel{\mathrm{def}}{=}&
\int  d{\vec \phi} \, e^{\frac{i}{2}z {\vec \phi}\cdot {\vec\phi} + \sum_{j=1}\frac{1}{j!} c_{j,d}
\left(-\frac{i}{2} {\vec\phi}\cdot {\vec\phi} \right)^j
-i g({\vec\phi})} \,
e^{ \frac{c_2}{4N}({\vec \phi} \cdot {\vec \phi})^2}
\qquad ,\qquad \qquad \nonumber\\
K({\vec \phi}_r,{\vec \phi}_s) & \stackrel{\mathrm{def}}{=}&
\frac{c_2}{2N}\left( {\vec \phi}_r \cdot {\vec \phi}_s \right)^2
\qquad \qquad \qquad
\label{hen.1aa}
\end{eqnarray}
The large $N$ behaviour of the functional integral is dominated by the saddle point function $g_{s.p.}$
  \begin{eqnarray}
&&\left( C\,g_{s.p.}\right)({\vec \psi})= \frac{N}{I[g_{s.p.}]}\left[
\frac{\delta}{\delta g({\vec \psi}) }I[g]\right]_{g=g_{s.p.}}=\frac{-i\,N}{I[g_{s.p.}]}
e^{\frac{i}{2}z {\vec \psi}\cdot {\vec\psi} + \sum_{j=1}\frac{1}{j!} c_{j,d}
\left(-\frac{i}{2} {\vec\psi}\cdot {\vec\psi} \right)^j
-i g_{s.p.}({\vec\psi})}\quad ,\qquad \nonumber\\
&& g_{s.p.}({\vec\psi})=-\frac{i\,N}{I[g_{s.p.}]}
\int d{\vec \phi}\,
 \frac{c_2}{2N}
 \left({\vec\phi}\cdot {\vec\psi} \right)^2\,
e^{\frac{i}{2}z {\vec \phi}\cdot {\vec\phi} + \sum_{j=1}\frac{1}{j!} c_{j,d}
\left(-\frac{i}{2} {\vec\phi}\cdot {\vec\phi} \right)^j
-i g_{s.p.}({\vec\phi})}
 \qquad \qquad
\label{hen.1ab}
\end{eqnarray}
The above equation shows that $g_{s.p.}( {\vec\psi})$ is a quadratic function of ${\vec\psi}$. We consider the ansatz  symmetric in the replica fields
\begin{eqnarray}
g_{s.p.}({\vec \phi})=\frac{1}{2}\beta(z, c_{j,d})\,{\vec \phi}\cdot {\vec \phi} 
 \qquad \qquad
\label{hen.1ac}
\end{eqnarray}
then the function $\beta(z,c_{j,d})$ is solution of the nonlinear equation
\begin{eqnarray}
&&\frac{\beta(z,c_{j,d})}{2}=-\frac{c_2}{n}\frac{1}{I(\beta)}\frac{\partial}{\partial z}I(\beta) \qquad \qquad {\rm where}
\nonumber\\
&&I(\beta)=\int d{\vec \phi}\,
e^{\frac{i}{2}(z-\beta) {\vec \phi}\cdot {\vec\phi} + \sum_{j=1}\frac{1}{j!} c_{j,d}
\left(-\frac{i}{2} {\vec\phi}\cdot {\vec\phi} \right)^j}
\qquad \qquad
\label{hen.1af}
\end{eqnarray}
This leads to the evaluation of the limiting resolvent
because
\begin{eqnarray}
\lim_{N\to \infty}g(z)=-2\lim_{n\to 0}\frac{1}{n}\frac{1}{I(\beta)}\frac{\partial}{\partial z}I(\beta)=
\frac{1}{c_2}\beta(z,c_{j,d}) 
\qquad \qquad
\label{hen.1ao}
\end{eqnarray}
and, by comparing $I(\beta)$ in eq.(\ref{hen.1af}) with the partition function $<Z_n(z)>$ of an ensemble of diagonal random matrices in eqs.(\ref{d.1}), (\ref{d.2}), we see that \footnote{If all diagonal cumulants vanish, $c_{j,d}=0$, the Gaussian integrals at the right side of eq.(\ref{hen.1af}) are easily evaluated and one obtains the quadratic equation $\beta^2(z)-z\beta(z)+c_2=0$. Together with eq.(\ref{hen.1ao}), one then recovers the semicircle law 
\begin{eqnarray}
\lim_{N\to \infty}g(z)=\frac{1}{c_2}\beta(z,c_{j,d}) =\frac{1}{2 c_2}\left(z-\sqrt{z^2-4c_2}\right) \qquad
\label{hen.1an}
\end{eqnarray}
}

\begin{eqnarray}
I(\beta)=\left[<Z_n(z-\beta)>\right]^{1/N} \qquad ,\qquad
\lim_{N\to \infty}g(z)=\lim_{N\to \infty}G_d\left(z-c_2 g(z)\right)
\qquad \qquad
\label{hen.1ag}
\end{eqnarray}
 where $G_d(z)$  
 is the resolvent of an ensemble of diagonal random matrices with probability distribution coded by the cumulants $c_{j,d}$.\\
 The functional equation for the limiting resolvents (\ref{hen.1ag})
 is the law of addition of random matrices proved by
 L. Pastur \cite{past}
\begin{eqnarray}
g(z)=\int d\lambda \frac{\rho_d(\lambda)}{z-c_2g(z)-\lambda} \qquad ,
\qquad \rho_d(\lambda)\stackrel{\mathrm{def}}{=}
-\frac{1}{\pi}{\rm Im}\,G_d(\lambda) \qquad \qquad
 \label{pa.2}
\end{eqnarray}

We remark that the result (\ref{pa.2}) holds unchanged if one considers the general case of non-Gaussian probability distribution for the off-diagonal matrix entries, provided Wigner scaling is still considered and the first cumulant vanishes, $c_{1,v}=0$. Then $c_{j,v}\to c_{j,v}/N^{j/2}$ for $j \geq 2$. Indeed, for a general probability distribution, the previous analysis is modified by the replacement
\begin{eqnarray}
\sum_{r,s}K \left({\vec \phi}_r,{\vec \phi}_s\right)=\sum_{r,s}\frac{c_2}{2N}
 \left({\vec \phi}_r,{\vec \phi}_s\right)^2 \to-\sum_{j=2}^{\infty}\frac{c_{j,v} }{j!N^{j/2}}\sum_{r,s}\left(-i{\vec \phi}_r\cdot {\vec \phi}_s\right)^j \qquad \qquad 
 \label{pa.3}
\end{eqnarray}
The series of new terms, inserted in eq.(\ref{hen.1ab})  only add contributions to the limiting resolvent, which are non-leading order in the large-$N$ limit.\\

\noindent {\bf The Wigner-rescaled random Laplacian matrices.}\\
The large-$N$ spectral distribution of a real symmetric Laplacian random matrix, where the matrix entries are Wigner rescaled, may be determined by the same saddle point method described above.\\
Let us first consider the simple case of Gaussian probability distribution. Then eq.(\ref{la.4}) becomes
\begin{eqnarray}
<Z_n(z)>=\int \prod_{i,\alpha} d\phi_i^{(\alpha)} \, e^{\frac{i}{2}z\sum_r ({\vec \phi}_r \cdot {\vec\phi}_r   
)-\frac{c_2}{16N}\sum_{r,s}\left[\left({\vec \phi}_r-{\vec \phi}_s\right)\cdot \left({\vec \phi}_r-{\vec \phi}_s\right)\right]^2}
\qquad \qquad 
 \label{pa.4}
\end{eqnarray}
Instead of eq.(\ref{hen.1aa}), we now have
 \begin{eqnarray}
<Z_n(z)>
&=&\sqrt{ \det C}\int  {\cal D}g \,e^{-\frac{1}{2} (g,Cg)+N \log \,I[g]}
 \qquad ,\qquad  \qquad \nonumber\\
I[g] & \stackrel{\mathrm{def}}{=}&
\int  d{\vec \phi} \, e^{\frac{i}{2}z {\vec \phi}\cdot {\vec\phi} 
-i g({\vec\phi})} \,\,
\qquad ,\qquad \qquad \nonumber\\
K({\vec \phi}_r,{\vec \phi}_s) & \stackrel{\mathrm{def}}{=}&
 \frac{c_2}{8N} \left[({\vec \phi}_r -{\vec \phi}_s)\cdot ({\vec \phi}_r-{\vec \phi}_s)\right]^2
\qquad \qquad \qquad
\label{pa.5}
\end{eqnarray}
The large $N$ behaviour of the functional integral is dominated by the saddle point function $g_{s.p.}$
  \begin{eqnarray}
&&\left( C\,g_{s.p.}\right)({\vec \psi})= \frac{N}{I[g_{s.p.}]}\left[
\frac{\delta}{\delta g({\vec \psi}) }I[g]\right]_{g=g_{s.p.}}=\frac{-i\,N}{I[g_{s.p.}]}
e^{\frac{i}{2}z {\vec \psi}\cdot {\vec\psi} 
-i g_{s.p.}({\vec\psi})}\quad ,\qquad \nonumber\\
&& g_{s.p.}({\vec\psi})=-\frac{i\,N}{I[g_{s.p.}]}
\int d{\vec \phi}\,
 \frac{c_2}{8N}\left[
 \left({\vec\phi}-{\vec \psi}\right)\cdot \left({\vec \phi}-{\vec\psi} \right)\right]^2\,
e^{\frac{i}{2}z {\vec \phi}\cdot {\vec\phi} -i g_{s.p.}({\vec\phi})}
 \qquad \qquad
\label{pa.6}
\end{eqnarray}
The above equation shows that $g_{s.p.}( {\vec\psi})$ is a quartic function of ${\vec\psi}$. We consider the ansatz,  symmetric in the replica fields, 
$$g_{s.p.}( {\vec\psi})=\alpha(z)+\frac{1}{2}\beta(z) {\vec \psi}\cdot {\vec \psi}+\gamma(z)\left({\vec \psi}\cdot {\vec \psi}\right)^2 $$ 
then $\gamma(z)=-ic_2/8$ and
the function $\beta(z)$ is the solution of the nonlinear equation
\begin{eqnarray}
&&\frac{\beta(z)}{2}=-\frac{c_2}{n}\left(1+\frac{n}{2}\right)
\frac{1}{ I(\beta)}
\frac{\partial}{\partial z}I(\beta) \qquad , \qquad {\rm where}
\nonumber\\
&&I(\beta)=
\int d{\vec \phi}\,\,
e^{\frac{i}{2}(z-\beta) {\vec \phi}\cdot {\vec\phi} -\frac{ c_2}{8}
\left( {\vec\phi}\cdot {\vec\phi} \right)^2}
 \qquad \qquad
\label{pa.7}
\end{eqnarray}
This leads to the evaluation of the limiting resolvent because eq.(\ref{hen.1ao}) still holds and
\begin{eqnarray}
\lim_{N\to \infty}g(z)=\lim_{N\to \infty}G_{d,g}\left(z-c_2 g(z)\right)
\qquad \qquad
\label{pa.8}
\end{eqnarray}
 where $G_{d,g}(z)$  
 is the resolvent of an ensemble of diagonal random matrices where the entries are i.i.d. centered Gaussian variables with variance  $c_2$.\\
The result (\ref{pa.8}) was derived in \cite{sta} by diagrammatic technique and Wick's theorem, to account for the Gaussian matrix entries with a constraint on each row of the matrix.\\

The present derivation shows that the result (\ref{pa.8}) for the limiting resolvent holds also in the general case where the off-diagonal entries of the real symmetric random Laplacian matrix are i.i.d. random variables with arbitrary probability distribution, coded by the cumulants $c_{j,v}$, provided they are Wigner rescaled and that the first cumulant vanishes, $c_{1,v}=0$.\\
Indeed, for a general probability distribution, the previous analysis is modified by the replacement
\begin{eqnarray}
&&\sum_{r,s}K \left({\vec \phi}_r,{\vec \phi}_s\right)=\sum_{r,s}\frac{c_2}{8N}\left[
 \left({\vec \phi}_r-{\vec \phi}_s\right)\cdot  \left({\vec \phi}_r-{\vec \phi}_s\right)\right]
^2 \to   \qquad \qquad \nonumber\\
&& \quad \to -\sum_{j=2}^{\infty}\frac{c_{j,v} }{j!N^{j/2}}\sum_{r,s}\left[
\frac{i}{2} \left({\vec \phi}_r- {\vec \phi}_s\right)\cdot
 \left({\vec \phi}_r- {\vec \phi}_s\right) \right]^j \qquad \qquad 
 \label{pa.9}
\end{eqnarray}
The series of new terms, inserted in eq.(\ref{pa.6})  only add contributions to the limiting resolvent, which are non-leading order in the large-$N$ limit.\\

Finally we remark that our result (\ref{hen.1ag}) refers to an ensemble where the diagonal elements are statistically independent from the off-diagonal entries. Then we merely derived, by a different technique a specific case of Pastur's law of addition of random matrices. The similar result about Laplacian matrices (\ref{pa.8}) is different because, at any finite $N$, the diagonal matrices are not statistically independent from the off-diagonal entries and Pastur theorem does not apply. Only at $N=\infty$, because of the central limit theorem, the diagonal entries become centered Gaussian random variables. Our saddle point analysis indicates that the rate of convergence for the diagonal entries to become statistically independent from the entries on the same row is fast enough to justify a naive use of Pastur's law also in this case.\\

Let $L$ be a real symmetric Laplacian random matrix where the off-diagonal matrix entries are i.i.d. random variables. The graphic-combinatorial method described in \cite{cic} may be used to evaluate $<{\rm Tr}\, L^k>$, for any $N$ and generic probability distribution. 
We  evaluated the Taylor expansion of the resolvent $g(z)$ in $1/z$ up to order $(1/z)^8$. Next it is trivial to rescale the results according to Wigner rescaling of the matrix entries, evaluate the large-$N$ limit and find results in agreement with eq.(\ref{pa.8}). This adds independent support to the saddle point analysis of this paper.\\

\noindent {\bf The random graph}\\
In models of sparse matrices or in the random graph model, the probability distribution for the entries of the random matrix $P(H_{ij})$ is such that the cumulants $c_j$ are of order $1/N$, in the large-$N$ limit, for every $j$. The general saddle point method of this section is still valid, but the non-linear integral equation obtained for $g_{s.p.}({\vec \psi})$ is more difficult.\\
In leading order the diagonal cumulants $c_{j,d}$ may be neglected and instead of eq.(\ref{hen.1ab}) we have
 \begin{eqnarray}
&& g_{s.p.}({\vec\psi})=-\frac{iN}{I[g_{s.p.}]}
\int d{\vec \phi} \,
e^{\frac{i}{2}z {\vec \phi}\cdot {\vec\phi} -i g_{s.p.}({\vec\phi})}
\sum_{j=1}^{\infty}
 \frac{c_{j,v}}{j!}
 \left(-i{\vec\phi}\cdot {\vec\psi} \right)^j\,
 \qquad ,\qquad \nonumber\\
&& I[g_{s.p.}]=\int d{\vec \phi} \,
e^{\frac{i}{2}z {\vec \phi}\cdot {\vec\phi} -i g_{s.p.}({\vec\phi})}
\qquad \qquad \qquad
\label{pa.10}
\end{eqnarray}

If one looks for solutions with spherical symmetry in the replica space, that is $g_{s.p.}({\vec\psi})=g_{s.p.}(|{\vec\psi}|)$,  the angular integrations are easily done and only the terms in the series (\ref{pa.10}) which are
proportional to $c_{j,v}$ with $j$ even integer contribute. We find

 \begin{eqnarray}
&&g_{s.p.}( |{\vec\psi}| )= \int_0^{\infty}\rho^{n-1}\,e^{\frac{i}{2}z \rho^2 -i g_{s.p.}(\rho)}
 F\left(\rho \, |{\vec\psi}|\right)d\rho \bigg/
\int_0^{\infty}\rho^{n-1}\,e^{\frac{i}{2}z \rho^2 -i g_{s.p.}(\rho)}
 d\rho \qquad ,\qquad \nonumber\\
&& F\left(\rho \, |{\vec\psi}|\right) =
\sum_{k=1}^{\infty}\frac{c_{2k}}{(2k)!}(-1)^k\rho^{2k}|{\vec\psi}|^{2k}a_{2k}\qquad ,\qquad \nonumber\\
&&a_{2k}=\frac{1\cdot 3\cdots (2k-3)(2k-1)}{n(n+2)\cdots (n+2k-4)(n+2k-2)} \qquad \qquad
\label{pa.11}
\end{eqnarray}

$${\rm For}\quad n\sim 0 \quad
\int_0^{\infty}\rho^{n-1}\,e^{\frac{i}{2}z \rho^2 -i g_{s.p.}(\rho)}\sim \frac{1}{n}
\qquad , \qquad a_{2k}\sim \left(\frac{1}{n}\right)\frac{3 \cdot 5\cdots (2k-3)(2k-1)}{2\cdot 4\cdots (2k-4)(2k-2)}\qquad$$
and
 we obtain a  simpler non-linear equation \footnote{As a simple check of our evaluation (\ref{pa.12}), one may insert $c_{2k}=p/N+O(1/N^2)$ and obtain $f\left( \rho \, |{\vec\psi}|\right) =-(p/N)|{\vec \psi}|J_1(
\rho \, |{\vec\psi}|)$ which reproduces the non-linear equation first derived by Rodgers and Bray \cite{rod}.} 

\begin{eqnarray}
&&g_{s.p.}( |{\vec\psi}| )= \int_0^{\infty} e^{\frac{i}{2}z \rho^2 -i g_{s.p.}(\rho)} f\left(\rho \, |{\vec\psi}|\right)d\rho 
\qquad ,\qquad \qquad\nonumber\\
&& f\left(\rho \, |{\vec\psi}|\right) =
\sum_{k=1}^{\infty}\frac{c_{2k}}{(k-1)!k!\,2^{2k-1}}(-1)^k\rho^{2k-1}|{\vec\psi}|^{2k}\qquad  \qquad \qquad
\label{pa.12}
\end{eqnarray}

A variety of approximate solutions may be found in \cite{rod} and \cite{seme}. An analogous non-linear integral equation is obtained for the random Laplacian. Its approximate solutions may be found in \cite{bra} , \cite{bir}, \cite{dea}, \cite{for}. A numerical solution is studied in \cite{pari}.\\

\section{Conclusion}
The limiting spectral distribution for  ensembles of real symmetric random Laplacian matrices with independent identically distributed Wigner-rescaled matrix entries has been derived in this paper. This is a generalization of previous results obtained with different techniques in the case of Gaussian matrix entries.\\

The replica method, combined with a Hubbard-Stratonovich functional identity ( reviewed and explained in Sect.3)
turns out to be a very powerful and general method in the study of random matrices. 
In the derivation of our new result, the Pastur law of addition of random matrices is recovered by the replica method.\\

Our derivation, up to the large-$N$ saddle point analysis, applies to Wigner rescaled matrix ensembles as well as to ensembles of sparse matrices , where usually $<v^k> \sim O(1/N)$. The relevance of the different scalings emerges at that point: in the case of ensembles of sparse matrices, the non-linear integral equation is very difficult to solve analytically.\\

{\bf Acknowledgments}\\
One of us (G.M.C.) is very grateful for useful and detailed discussions with Yan Fyodorov and Luca Molinari.\\

\end{document}